\documentclass[namedreferences]{SolarPhysics}
\usepackage[optionalrh]{spr-sola-addons} % For Solar Physics
\usepackage{graphicx}        % For eps figures, newer & more powerfull
\usepackage{color}           % For color text: \color command
\usepackage{url}             % For breaking URLs easily trough lines
\newcommand{\etal}{{\it et al.}}

\newcommand{\aap}{    {\it Astron. Astrophys.}}

\newcommand{\apj}{    {\it Astrophys. J.}}

\newcommand{\nat}{    {\it Nature}}

\newcommand{\pasj}{   {\it Pub. Astron. Soc. Japan}}

\newcommand{\solphys}{{\it Solar Phys.}}

\newcommand{\ssr}{    {\it Space Sci. Rev.}}

\begin{document}

\begin{article}

\begin{opening}

\title{Quasi periodic oscillations of solar active regions in connection with their flare activity
-- NoRH observations}

\author{V.E.~\surname{Abramov-Maximov}$^{1}$\sep
        G.B.~\surname{Gelfreikh}$^{1}$\sep
        K.~\surname{Shibasaki}$^{2}$
       }
\runningauthor{V.E.~Abramov-Maximov \etal} \runningtitle{Quasi
periodic oscillations and flare activity}

\institute{$^{1}$ Central Astronomical Observatory at Pulkovo,
Russian Acad. Sci., St.Petersburg, 196140, Russia\\
email: \url{beam@gao.spb.ru}\\
$^{2}$ Nobeyama Solar Radio Observatory, Minamisaku, Nagano, Japan\\
email: \url{shibasaki@nro.nao.ac.jp} \\}

\begin{abstract}

The sunspot-associated sources at the frequency of 17~GHz give
information on plasma parameters in the regions of magnetic field
about $B=2000$~G at the level of the chromosphere-corona
transition region. The observations of short period (from 1 to
10~minutes) oscillations in sunspots reflect propagation of
magnetohydrodynamic (MHD) waves in the magnetic flux tubes of the
sunspots. We investigate the oscillation parameters in active
regions in connection with their flare activity. We confirm the
existence of a link between the oscillation spectrum and flare
activity. We find differences in the oscillations between
pre-flare and post-flare phases. In particular, we demonstrate a
case of powerful three-minute oscillations that start just before
the burst. This event is similar to the cases of the precursors
investigated by Sych,~R.~\textit{et al.}~(\textit{Astron.
Astrophys.} \textbf{505}, 791, 2009). We also found well-defined
eight-minute oscillations of microwave emission from sunspot. We
interpret our observations in terms of a relationship between MHD
waves propagating from sunspot and flare processes.

\end{abstract}
\end{opening}
%-------------------------------------------------

\section{Introduction}
\label{S-Introduction}

Quasi-periodic oscillations (QPO) are registered practically in
all wavelength ranges and in all structures of the solar
atmosphere~(Lites \textit{et al.}, 1998; Bogdan, 2000; Fludra,
2001; Bogdan and Judge, 2006; Kosovichev, 2009).
%\cite{Lites98,Bogdan00,Fludra01,Bogdan06,Kosovichev09}.
The periods of oscillations typically range from seconds to hours
and perhaps days. The dominant periods of the oscillations are
three and five minutes. Most of these oscillations are of
intermittent nature: both their amplitude and frequency vary with
time. The oscillations are often visible as trains of a few
periods. The oscillations with different periods have different
physical nature. The short-period oscillations are caused by
running waves. Their intermittency is probably caused by
instability of physical conditions in the regions of generation
and propagation.

Studying oscillations can help us to understand  such fundamental
astrophysical problems as accumulation and release of energy,
physics of the coronal heating and the origin of flares. The
investigation of oscillation processes in the solar corona is a
powerful tool for coronal plasma diagnostics~(Nakariakov and
Erdelyi, 2009).%\cite{Nakariakov09}.

Observations of QPOs of the solar radio emission have been carried
out for more than 40 years~\cite{Durasova71}. The first
observations showed that oscillations are mostly produced in solar
active regions and their parameters reflect the development of
flare activity~\cite{Kobrin73,Aleshin73}. However, in these
studies the spatial and time resolution of radio telescopes was
not adequate.

Significant progress in this field was achieved with the help of
large modern radio telescopes. Especially rich information was
gained from the Nobeyama Radioheliograph (NoRH)~\cite{Nakajima94}.
The NoRH has been observing the Sun since 1992. The observations
are carried out every day (7\,--\,8 hours daily) with spatial
resolution of 10\,--\,20~arcsec and cadence of 1 sec.

The most sensitive analysis of the radio oscillations reflecting
the  plasma resonance phenomena for MHD oscillations was found for
the sunspot associated sources~(Gelfreikh~\textit{et al.}, 1999;
Shibasaki, 2001; Nindos~\textit{et al.}, 2002;
Gelfreikh~\textit{et al.}, 2006).
%\cite{Gelfreikh99,Shibasaki01,Nindos02,Gelfreikh06}.
Such oscillations reflect the oscillations of the plasma
parameters (temperature and magnetic field) at the level of the
chromosphere-corona transition region (CCTR), where the strength
of the magnetic field is about $B=2000$~G. The analysis has shown
the presence of regular oscillations of the radio brightness
(registered both in $I$ and $V$ Stokes parameters). The periods of
oscillations vary from fractions of a minute to several hours. In
reality they have a different nature. The nature of oscillations
is still in the stage of discussion. However, some mechanisms have
been identified. The three- and five-minute oscillations in
sunspots are well known from optical
observations~\cite{Uchida75,Horn97} and also analyzed in
space-borne data of EUV lines~(Brynildsen~\textit{et al.}, 1999a;
Brynildsen~\textit{et al.}, 1999b; De~Moortel~\textit{et al.},
2002; Brynildsen~\textit{et al.}, 2003; King~\textit{et al.},
2003).
%\cite{Brynildsen99a,Brynildsen99b,De_Moortel02,Brynildsen03,King03}.
It was shown that they usually reflect propagation of MHD waves
from the chromosphere towards corona. In some cases the waves
propagate along coronal loops rooted in
sunspots~%(De~Moortel~\textit{et al.}, 2002).
\cite{De_Moortel02}.

Recently \inlinecite{Sych09} established the relationship between
three-minute oscillations of the microwave emission and
quasi-periodic pulsations (QPP) in flares. They found two cases of
increase in the power of the three-minute oscillations just before
the flares. The oscillations are interpreted as slow
magnetoacoustic waves which can cause the QPP.

In many sunspot-associated sources longer periods of QPO (dozens
of minutes) are also found. These periods are typical for the
coronal loops
oscillations~\cite{Gelfreikh06,Bakunina09,Chorley10}. Some of
these may begin in a sunspot region reflecting possibly acoustic
oscillations in coronal loops (see also~\opencite{Nakariakov04}).
Much longer periods, up to several hours, are also observed for
most of the discussed radio sources. Similar long periods were
found from the optical observations of sunspots~(Efremov,
Parfinenko, and Solov'ev, 2010).
%\cite{Efremov10}.
Global oscillations of the Sun may also be reflected in some cases
of sunspot radio emission.

In summary, study of QPO observations of  sunspot-associated
sources may lead to significant progress in understanding of
physical processes in solar plasma. Such observations also contain
information on MHD wave propagation in the solar atmosphere.
Although so far we have not reached satisfactory interpretation of
the physical nature of all observed types of the discussed QPO,
there is no doubt that the detailed analysis of their parameters
brings new insight concerning the plasma structures in the studied
active regions.

The main aim of this work is to study the difference in
oscillations observed in pre-flare and post-flare phases and to
investigate the reconstruction of the magnetic structure of ARs.

\section{Observations}
\label{S-Observations}

\subsection{Methods}
\label{S-Methods}

In this study we used  the NoRH daily (7\,--\,8~hours every day)
observations at 1.76~cm~(17~GHz). The radio maps of the whole
solar disk were synthesized with cadence of ten seconds and ten
seconds averaging. The spatial resolution of the radio maps is
about 10\,--\,15 arcsec.

NoRH provides intensity and circular polarization (I and V,
respectively) solar images. At 17 GHz the total intensity and
polarization data usually yield similar results for the sunspot
associated-sources and this is also the case for our study. This
behavior is due to the cyclotron nature of emission. The polarized
emission at 1.76 cm is generated at the third harmonic of electron
gyrofrequency in a layer where the magnetic field is about
$B=2000$~G~\cite{Akhmedov82,Shibasaki94,Nindos00}. The pioneer
study of oscillations using NoRH data~\cite{Gelfreikh99} was based
on polarization data only. \inlinecite{Shibasaki01} used time
series of the peak brightness temperature of the intensity.
Additionally, he used correlation plots for the right and the left
circular polarization, but he did not use solar images in
polarization. Some authors
\cite{Gelfreikh06,Chorley10,Dzhimbeeva11} used total intensity
data only. In this study we only present the intensity data.

Figures~\ref{F-2001-09-11}b,\ref{F-2002-03-14}b,\ref{F-2002-10-07}b
show full disk images and magnified images of the investigated
sources as well. All radio sources are compact. The structure of
the source in the Figure~\ref{F-2002-03-14}b is more complicated.
But there is a compact feature in the left bottom part of the
magnified image. This feature is brighter. The size of the feature
is comparable to the synthesized beam size. So, we can consider
all sources as unresolved and use the peak brightness temperature
to generate time series.

{\samepage The basic steps of data processing are:
\begin{itemize}
    \item interactive extraction of the selected region-of-interest (ROI) from the initial
    image,
    \item computation of the new position of the ROI corrected for
solar rotation for each image according to the time of the
observation,
    \item extraction of the ROI from
    all the images,
    \item computation of the time series of the maximum brightness
temperature over each ROI,
    \item spectral wavelet analysis of the time profiles.
\end{itemize}}

For the analysis of the time profiles we used wavelet spectra with
Morlet functions of the sixth order as a base
function~\cite{Torrence98}.

\subsection{The Active Region 9608 on 11 September 2001}
\label{S-AR9608}

\begin{figure}
   \centerline{\hspace*{0.05\textwidth}
               \includegraphics[width=0.388\textwidth,clip=]{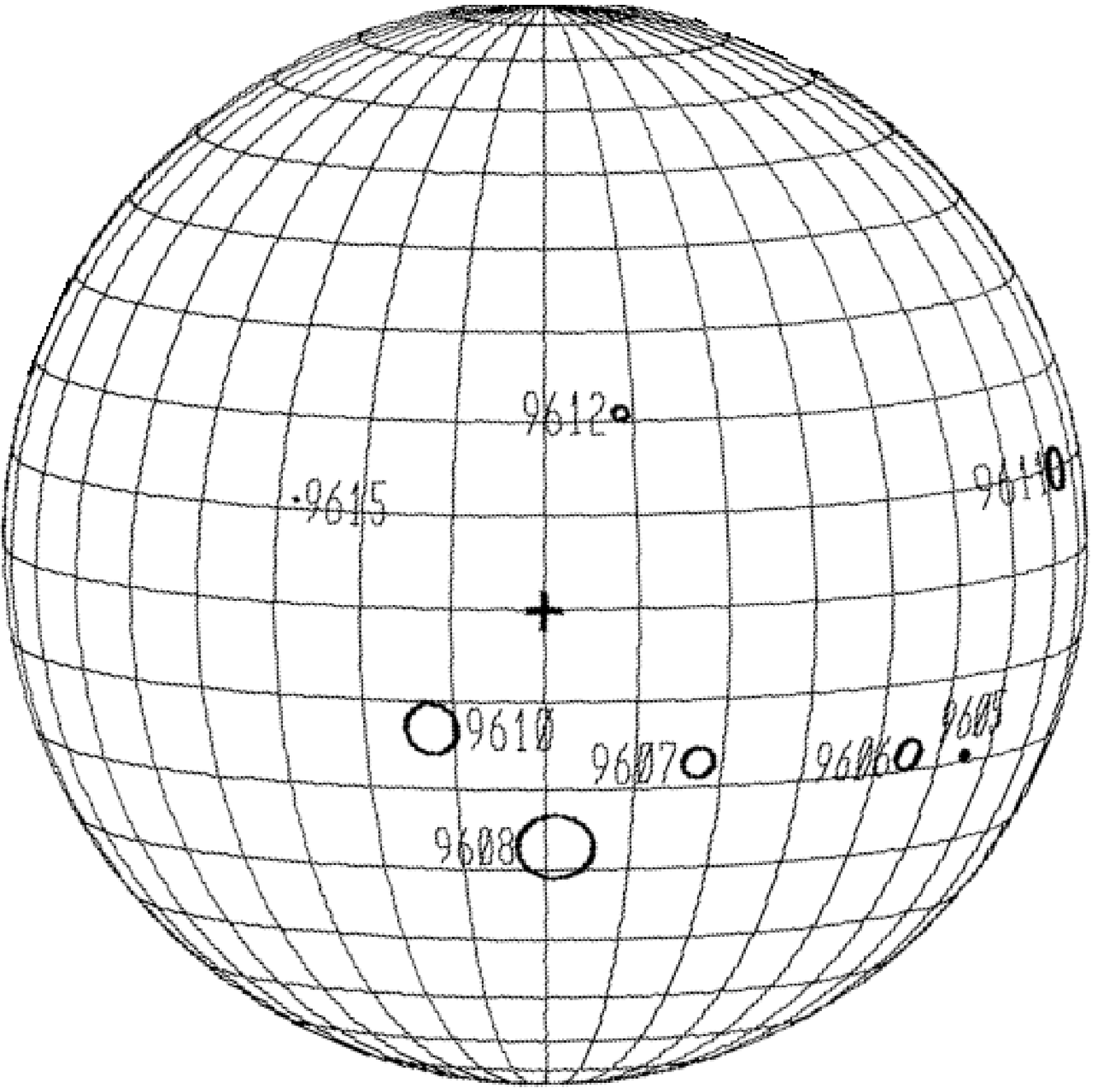}
               \hspace*{0.1\textwidth}
               \includegraphics[width=0.554\textwidth,clip=]{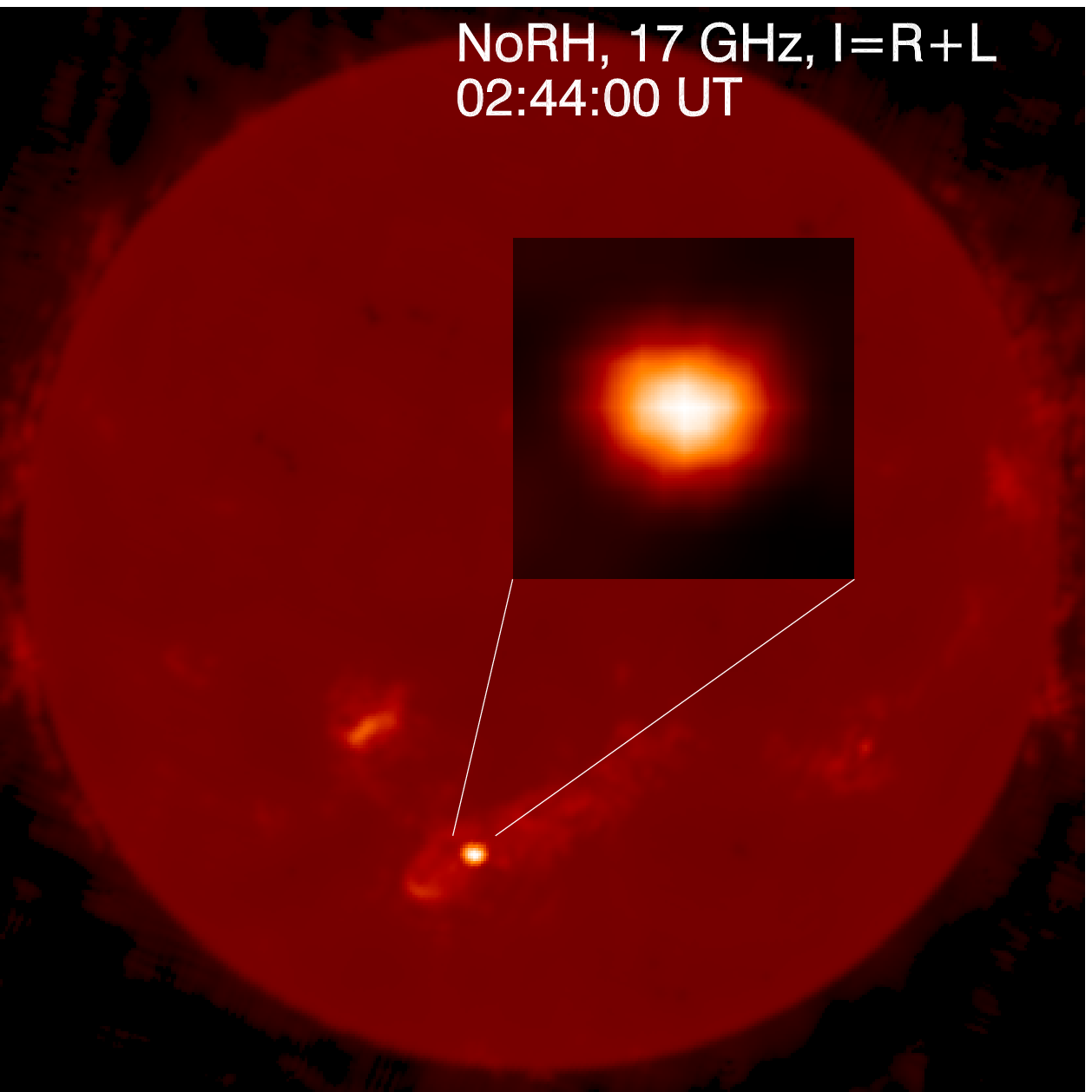}}
   \vspace{-0.38\textwidth}
   \centerline{\Large \bf
               \hspace{0.01\textwidth}\color{black}{a}
               \hspace{0.45\textwidth} \color{white}{b}
   \hfill}
   \vspace{0.38\textwidth}
   \vspace{0.02\textwidth}
   \centerline{\hspace*{-0.05\textwidth}
               \includegraphics[width=0.6\textwidth,clip=]{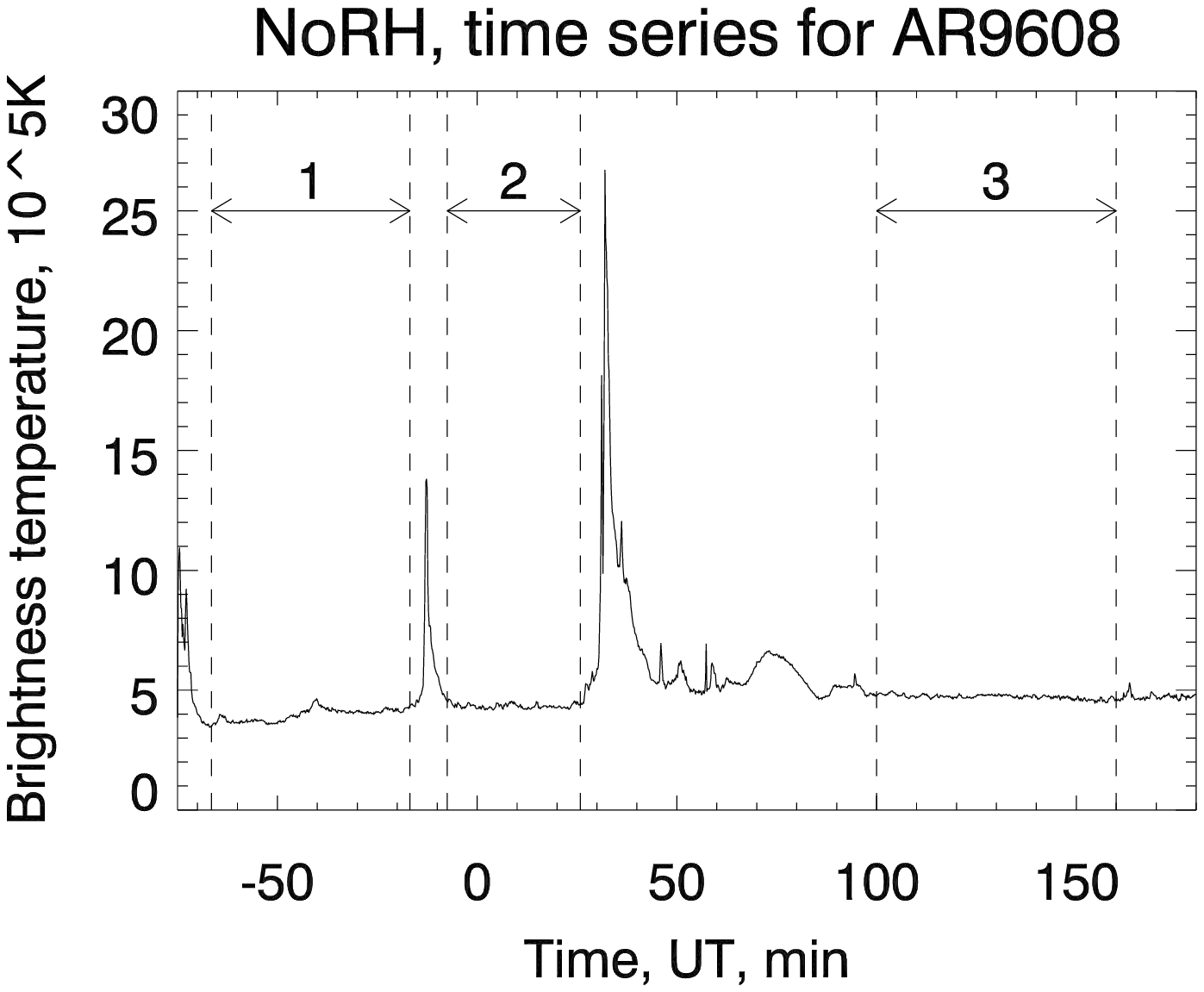}
               \hspace{-0.15\textwidth}
               \includegraphics[width=0.6\textwidth,clip=]{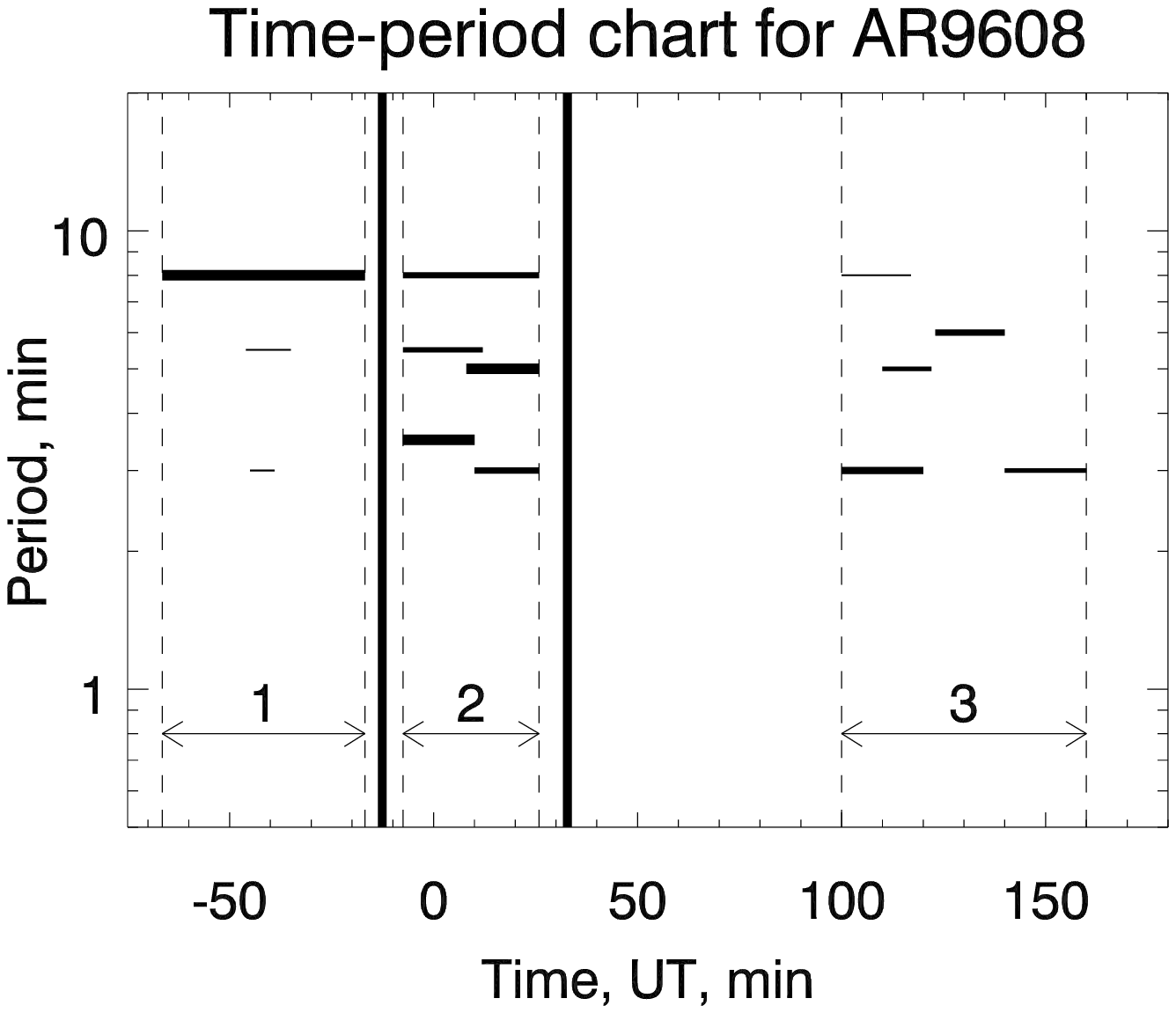}}
   \vspace{-0.46\textwidth}
   \centerline{\Large \bf
               \hspace{0.03\textwidth} \color{black}{(c)}
               \hspace{0.4\textwidth} \color{black}{(d)}
   \hfill}
   \vspace{0.46\textwidth}
\caption{NOAA~9608, 11~September 2001. Top: (a) active region
position on the solar disk (Mees Solar Observatory, 16:32~UT), (b)
radio map at 17~GHz (2:44~UT) and enlarged ROI, the size of the
ROI is about $70''\times70''$. Bottom: (c) time series of the peak
brightness temperature at 17~GHz, vertical dashed lines show three
time intervals (before the first burst, between the bursts, after
the second burst) of the analysis, (d) time-period chart showing
the life-times of the significant oscillations by the horizontal
lines, the thickness of the lines reflects the wavelet power, the
thick vertical lines show the bursts.} \label{F-2001-09-11}
\end{figure}

Figure~\ref{F-2001-09-11}a shows the position of the AR~9608. It
was situated at the central meridian but shifted by half a solar
radius toward the south pole. The NoRH radio map is presented in
Figure~\ref{F-2001-09-11}b. It is clearly visible that the
analyzed AR has the largest sunspot and the brightest radio source
on the solar disk. The parameters of the source are in good
agreement with a gyro-resonance radio source at the level of the
low solar corona with the field $B=2000$~G. Two large bursts have
been registered during the four-hour interval of analysis
(Figure~\ref{F-2001-09-11}c). So, for analysis of the variations
of the spectra in quiet state we should analyze three intervals.
These are rather short, and analysis of long periods (tens of
minutes e.g.) is not possible.

\begin{figure}
   \centerline{%\hspace*{0.05\textwidth}
               \includegraphics[width=0.5\textwidth,clip=]{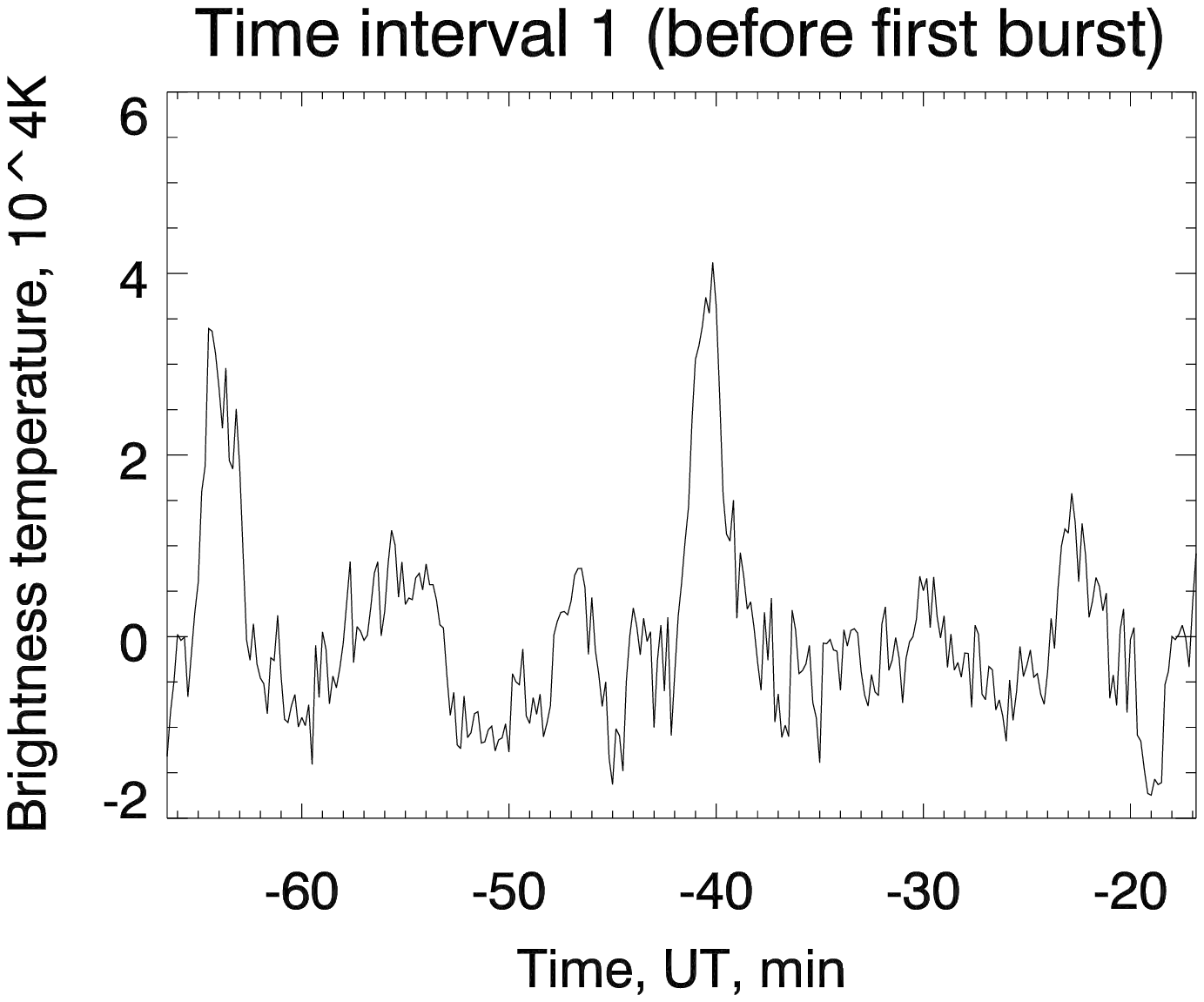}
%               \hspace*{0.1\textwidth}
               \includegraphics[width=0.5\textwidth,clip=]{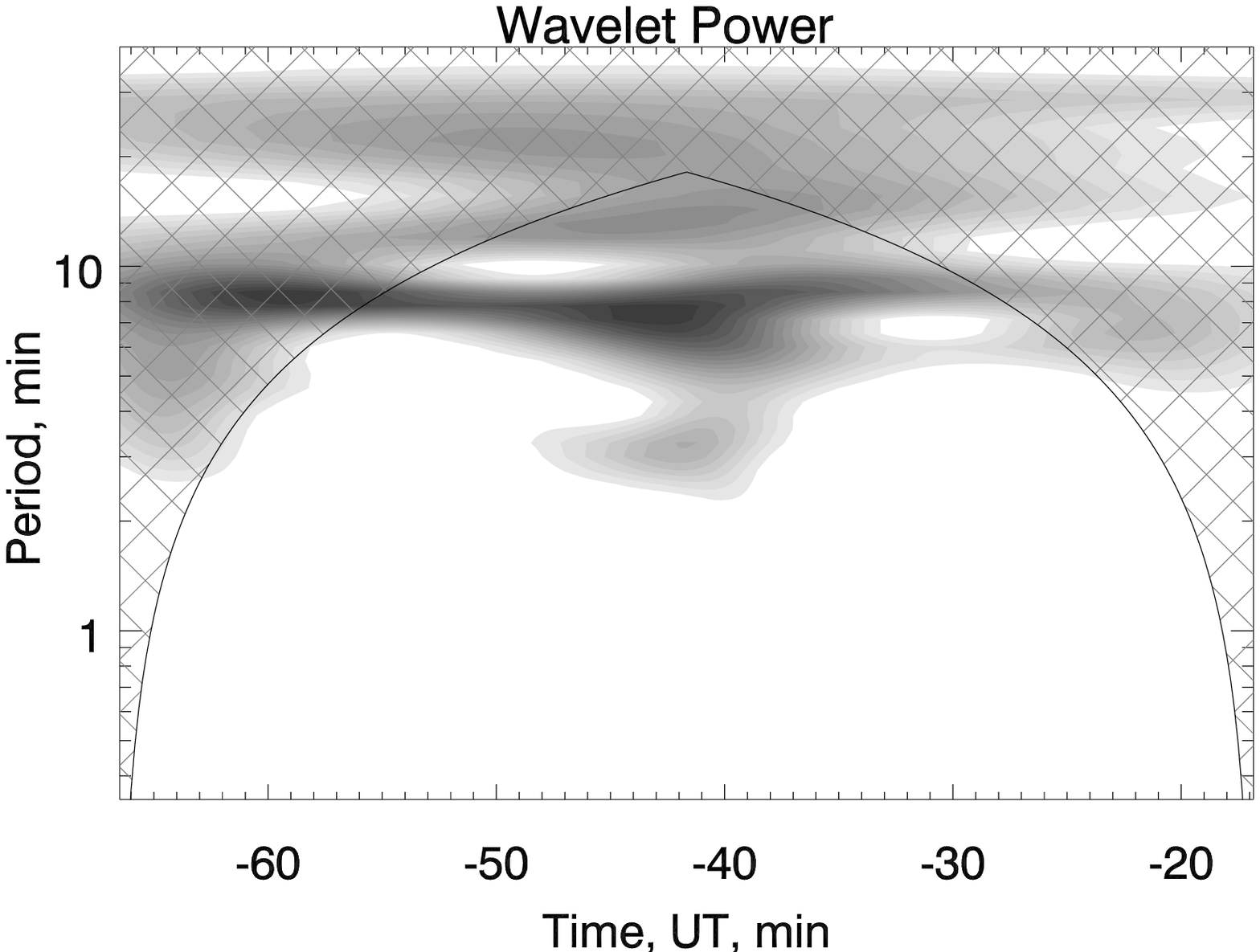}}
   \vspace{-0.38\textwidth}
   \centerline{\Large \bf
               \hspace{0.01\textwidth}\color{black}{a}
               \hspace{0.45\textwidth} \color{black}{b}
   \hfill}
   \vspace{0.38\textwidth}
   \centerline{\includegraphics[width=0.5\textwidth,clip=]{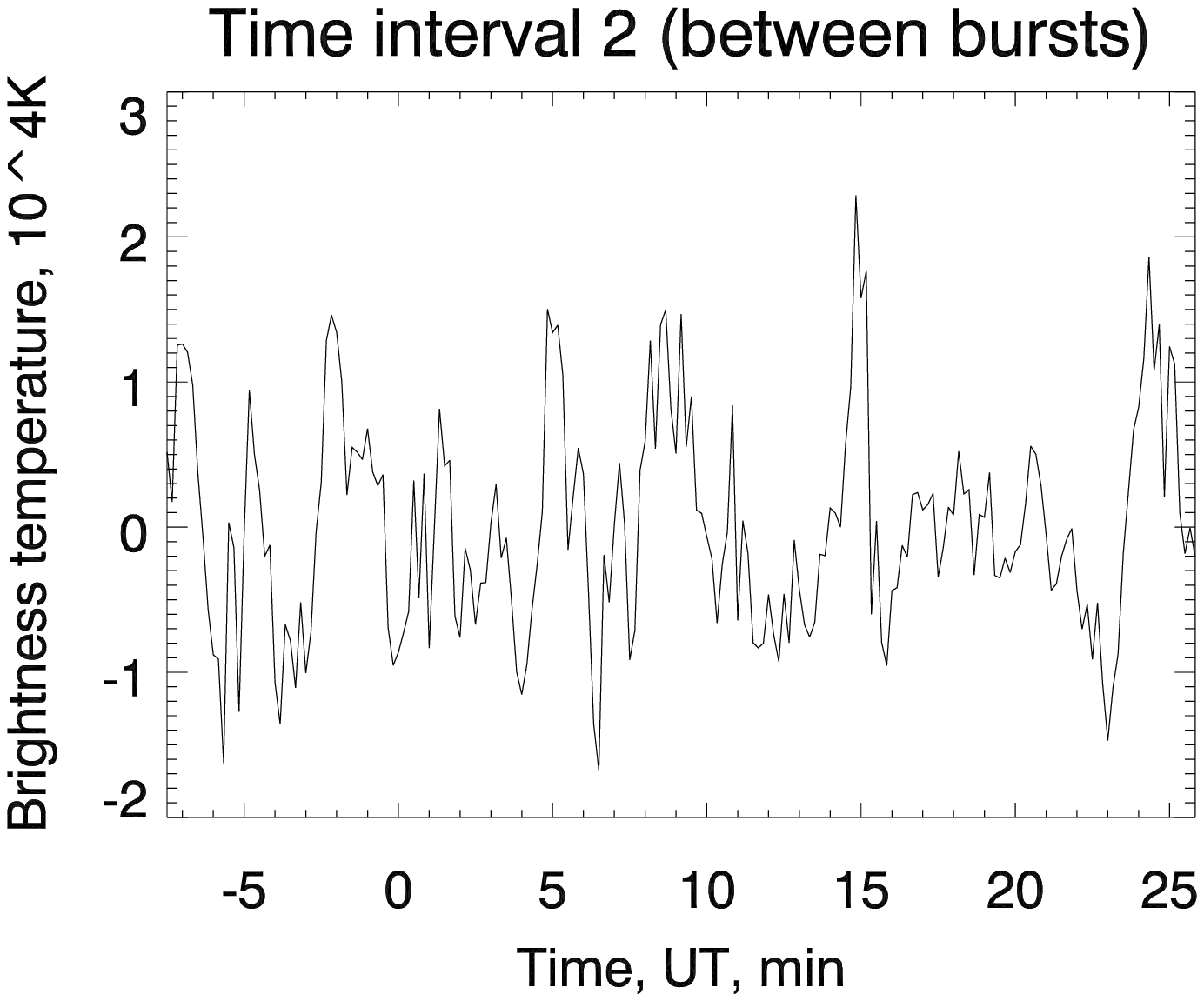}
%               \hspace*{0.1\textwidth}
               \includegraphics[width=0.5\textwidth,clip=]{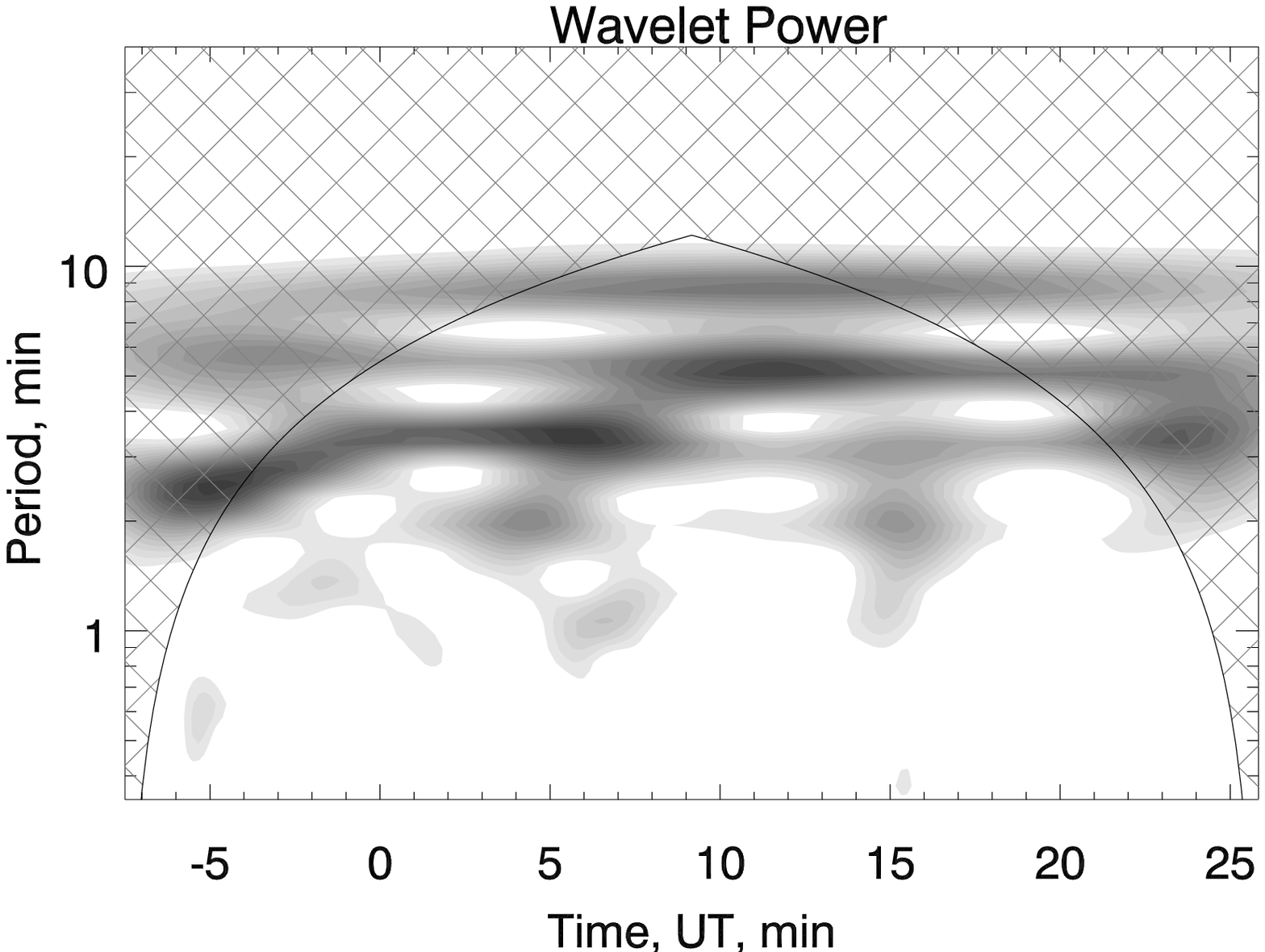}}
   \vspace{-0.39\textwidth}
   \centerline{\Large \bf
               \hspace{0.01 \textwidth} \color{black}{c}
               \hspace{0.45\textwidth} \color{black}{d}
   \hfill}
   \vspace{0.39\textwidth}
   \centerline{\includegraphics[width=0.5\textwidth,clip=]{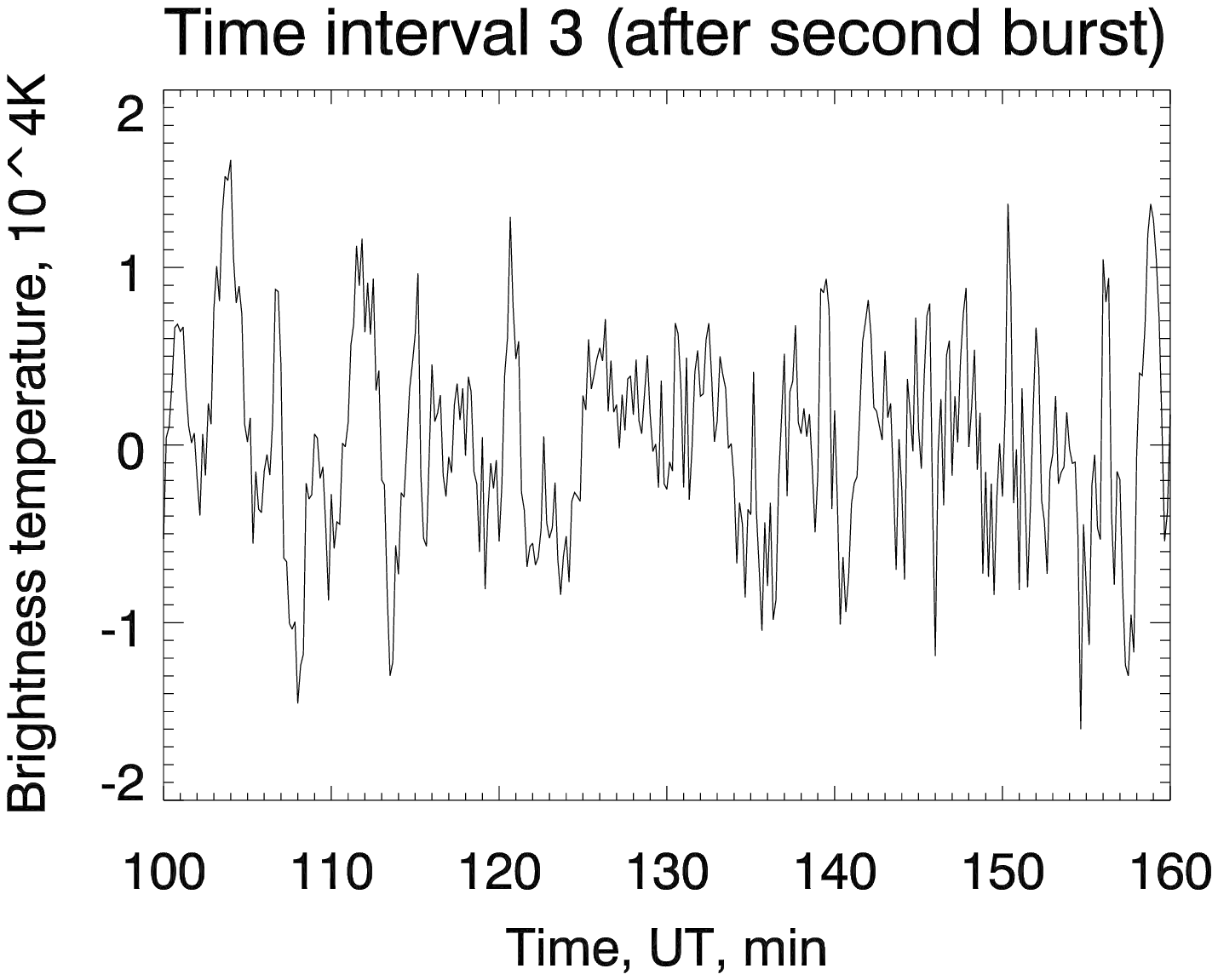}
%               \hspace*{0.1\textwidth}
               \includegraphics[width=0.5\textwidth,clip=]{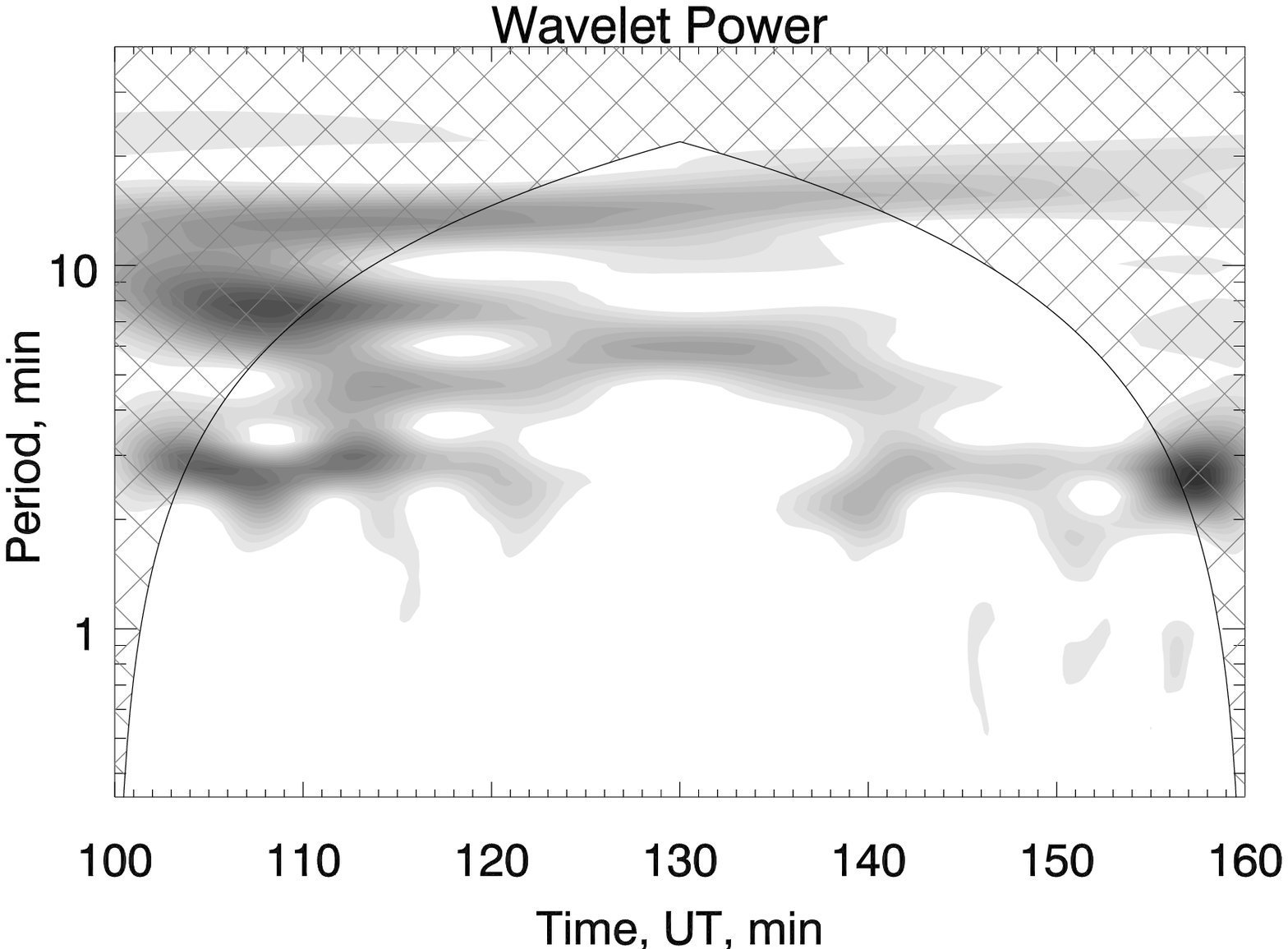}}
   \vspace{-0.39\textwidth}
   \centerline{\Large \bf
               \hspace{0.01 \textwidth} \color{black}{e}
               \hspace{0.45\textwidth} \color{black}{f}
   \hfill}
   \vspace{0.39\textwidth}

\caption{NOAA~9608, 11~September 2001. Top: (a) first time
interval of the time series (the trend has been subtracted)  of
the peak brightness temperature at 17~GHz, (b) wavelet spectrum of
the first time interval of the time series, darker regions
correspond to higher power, the crosshatched area shows the cone
of influence (COI). Middle: (c) second time interval of the time
series (the trend has been subtracted)  of the peak brightness
temperature at 17~GHz, (d) wavelet spectrum of the second time
interval of the time series. Bottom: (e) third time interval of
the time series (the trend has been subtracted) of the peak
brightness temperature at 17~GHz, (f) wavelet spectrum of the
third time interval of the time series.} \label{F-AR9608}
\end{figure}

Figure~\ref{F-AR9608} depicts the time intervals of the
trend-subtracted time series (a, c, and e) and corresponding
wavelet spectra (b, d, and f). To summarize information on
oscillations we have made time-period charts based on the wavelet
spectra. Figure~\ref{F-2001-09-11}d shows the time-period chart
for AR~9608. We register short wave-train of weak three-minute
oscillations during the first time interval
(Figures~\ref{F-2001-09-11}d and~\ref{F-AR9608}b). Probably
five-minute oscillations are present. We can clearly see the
oscillations with period of nearly eight minutes during the whole
interval. During the time interval between the bursts
(Figures~\ref{F-2001-09-11}d and~\ref{F-AR9608}d) the three- and
five-minute oscillations are present. The eight-minute
oscillations are seen also, but their power is weaker. After the
second burst (Figures~\ref{F-2001-09-11}d and~\ref{F-AR9608}f)
there are unstable wave-trains of the three- and five-minute
oscillations and the eight-minute oscillations are practically
absent.

It is known that the three-minute oscillations are a typical
feature of the sunspot atmosphere (not at all heights, however).
So, the appearance or disappearance of oscillations in radio
wavelengths reflect the reconstruction of the structure of
magnetic flux tube of the large sunspot at the level of CCTR (or
lower corona).

\subsection{The Active Region 9866 on 14 March 2002}
\label{S-AR9866}

\begin{figure}
   \centerline{\hspace*{0.05\textwidth}
               \includegraphics[width=0.388\textwidth,clip=]{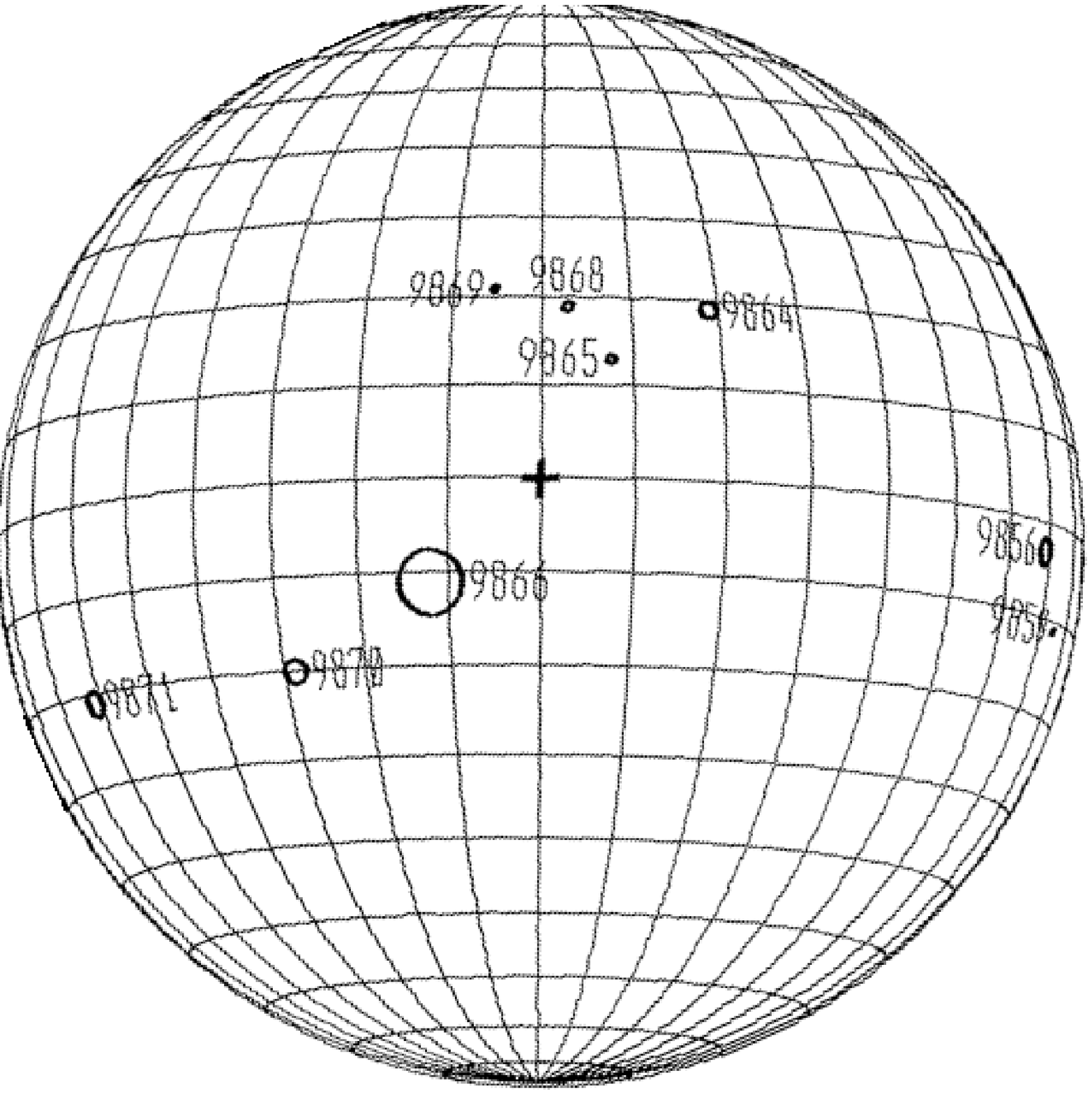}
               \hspace*{0.1\textwidth}
               \includegraphics[width=0.554\textwidth,clip=]{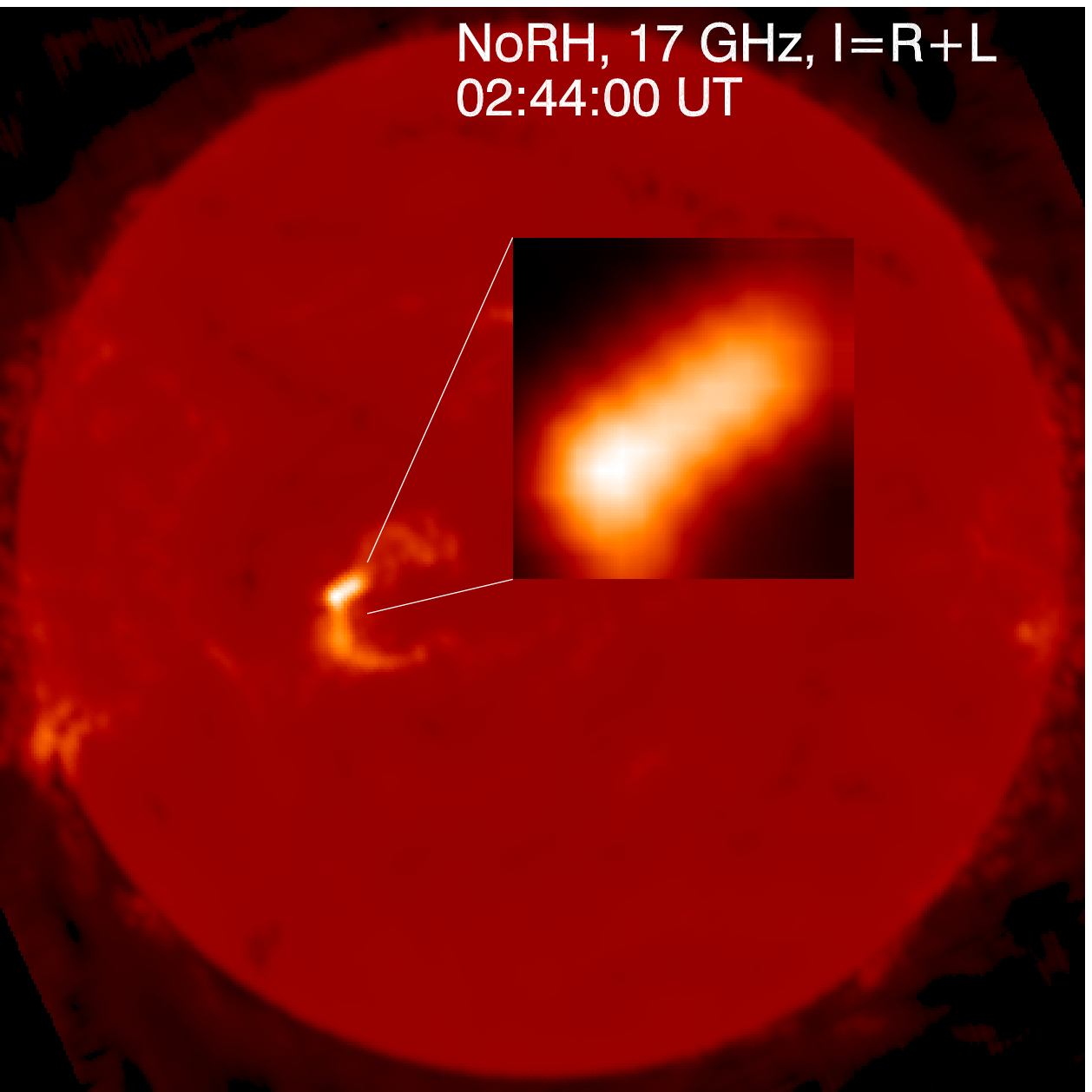}}
   \vspace{-0.38\textwidth}
   \centerline{\Large \bf
               \hspace{0.01\textwidth}\color{black}{a}
               \hspace{0.45\textwidth} \color{white}{b}
   \hfill}
   \vspace{0.38\textwidth}
   \vspace{0.02\textwidth}
   \centerline{\hspace*{-0.05\textwidth}
               \includegraphics[width=0.6\textwidth,clip=]{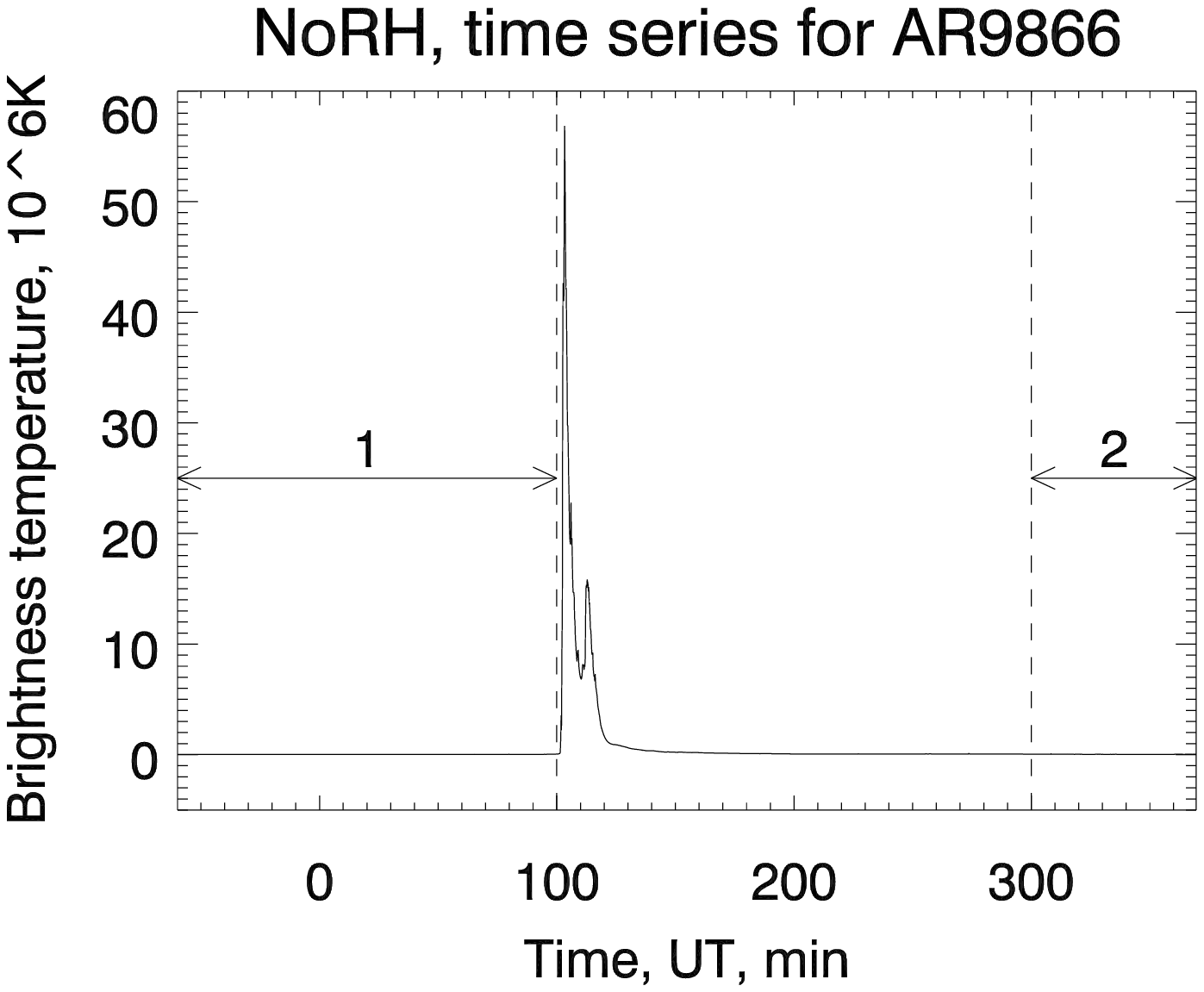}
               \hspace{-0.15\textwidth}
               \includegraphics[width=0.6\textwidth,clip=]{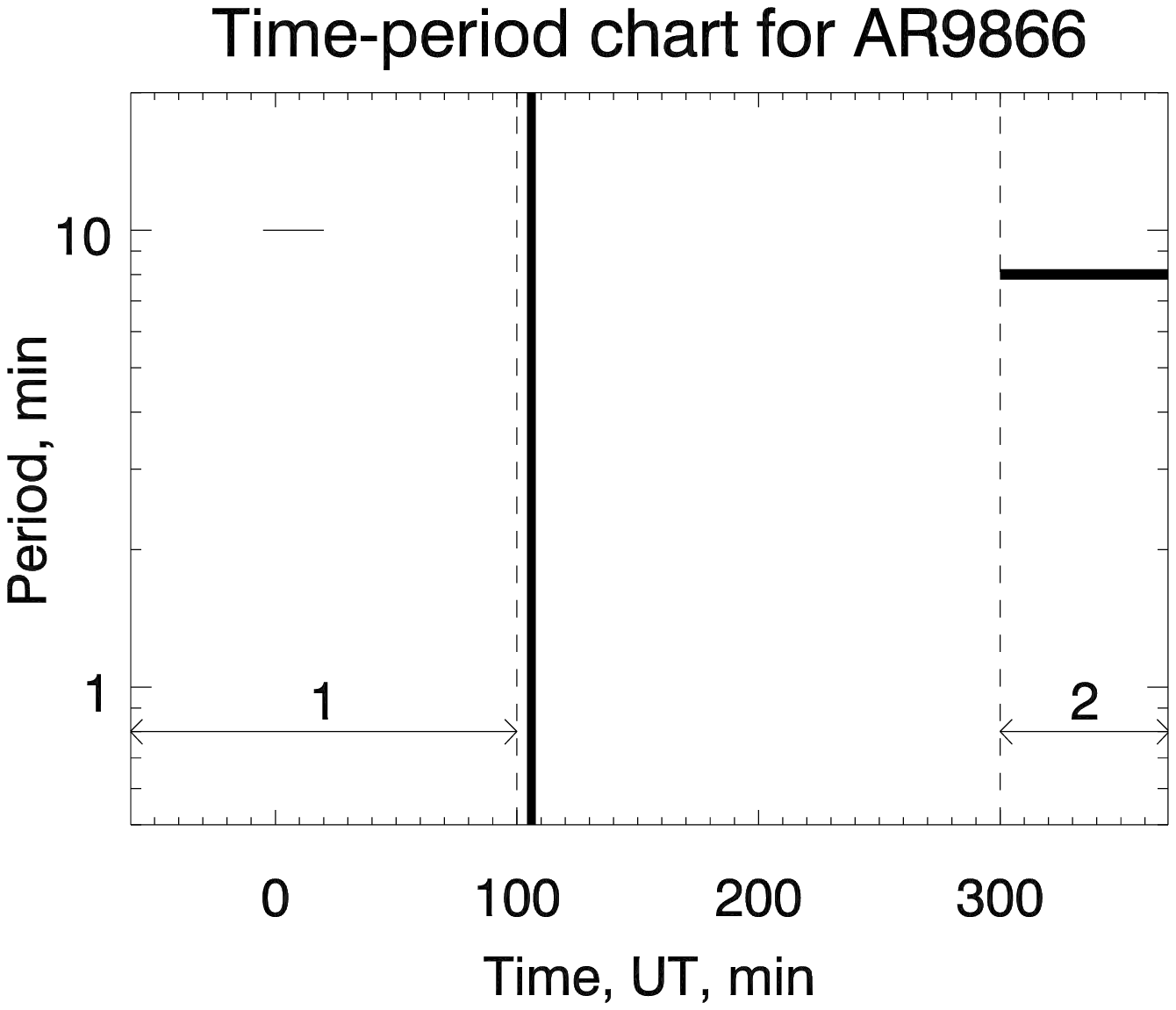}}
   \vspace{-0.46\textwidth}
   \centerline{\Large \bf
               \hspace{0.03\textwidth} \color{black}{(c)}
               \hspace{0.4\textwidth} \color{black}{(d)}
   \hfill}
   \vspace{0.46\textwidth}
\caption{NOAA~9866, 14~March 2002. Top: (a) active region position
on the solar disk (Mees Solar Observatory, 16:32~UT), (b) radio
map at 17~GHz (2:44~UT) and enlarged ROI, the size of the ROI is
about $90''\times90''$. Bottom: (c) time series of the peak
brightness temperature at 17~GHz, vertical dashed lines show two
time intervals (before the burst and after the burst) for the
analysis, (d) time-period chart showing the life-times of the
significant oscillations by the horizontal lines, the thickness of
the lines reflects the wavelet power, the thick vertical line
shows the bursts.} \label{F-2002-03-14}
\end{figure}

This active region includes a large bipolar sunspot (see
Figure~\ref{F-2002-03-14}(a,b)), crossing the central meridian a
day before. One large burst was registered at 1:40~UT during
5.3--hour period of the observations~(Figure~\ref{F-2002-03-14}c).
The flare happened in the west part of the region, which had a
bipolar magnetic structure. Generating the time series, we
verified the positions of the peak (X,Y-coordinates of the maximum
brightness temperature) to make sure that the peak belongs to the
same area. Actually, the peak position varied during the flare
because the flare had a complicated structure. However, we did not
analyze any details of the time series during the flare. We used
time series before and after flare, when the source was stable.

The time-period chart is shown in Figure~\ref{F-2002-03-14}d. One
can notice the appearance of powerful eight-minute oscillations
after the burst.

\begin{figure}
   \centerline{\includegraphics[width=0.5\textwidth,clip=]{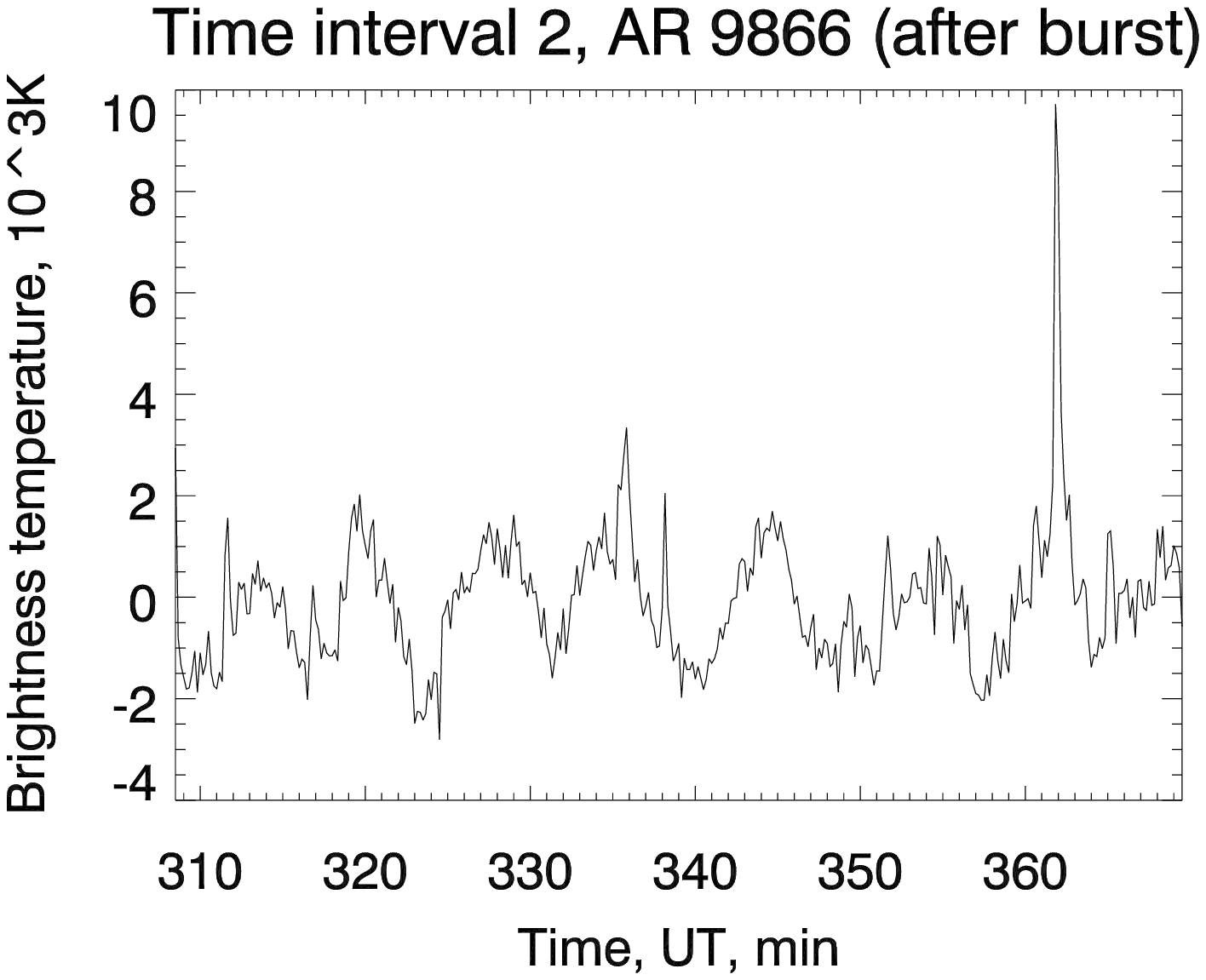}
%               \hspace*{0.1\textwidth}
               \includegraphics[width=0.5\textwidth,clip=]{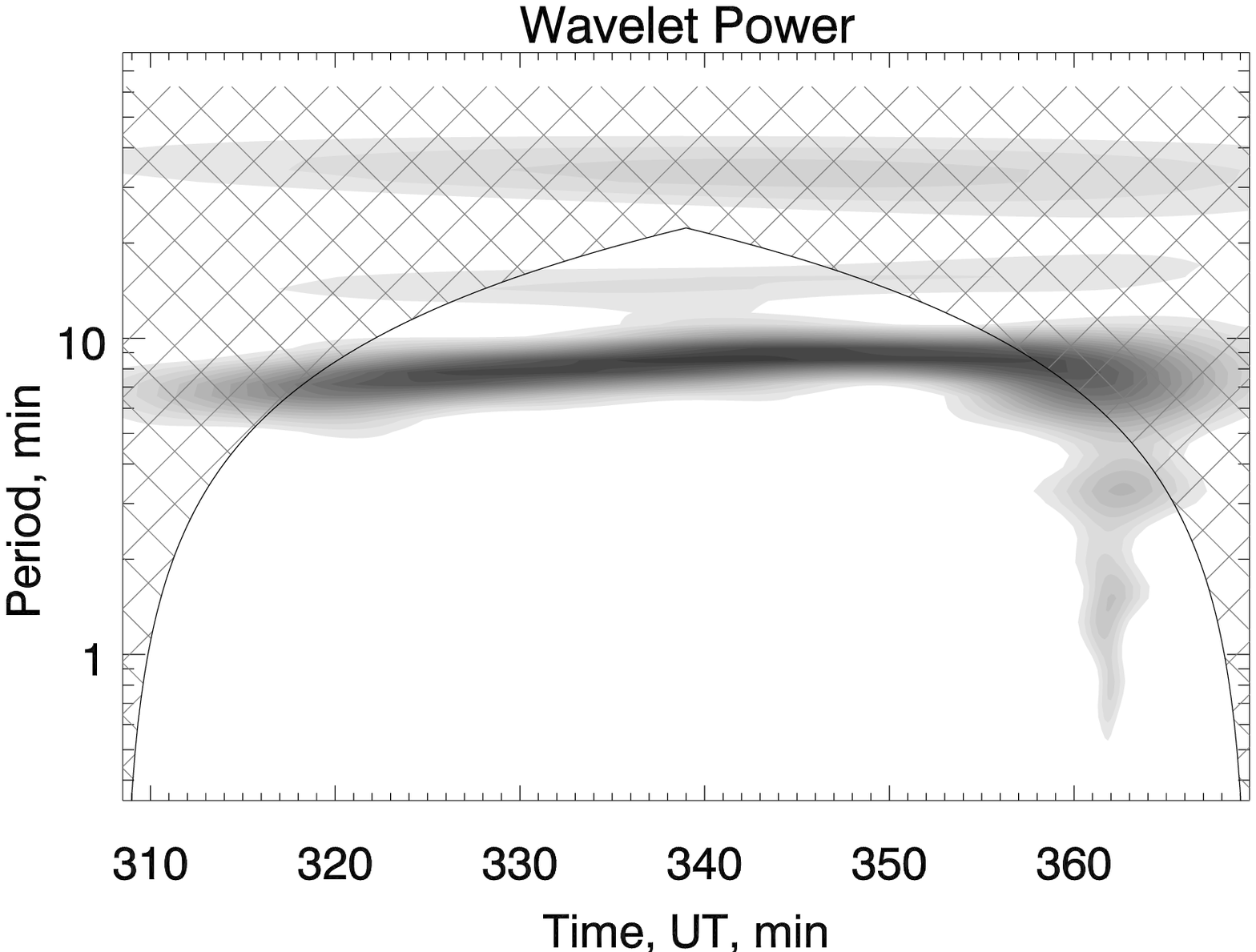}}
   \vspace{-0.38\textwidth}
   \centerline{\Large \bf
               \hspace{0.01 \textwidth} \color{black}{a}
               \hspace{0.45\textwidth} \color{black}{b}
   \hfill}
   \vspace{0.38\textwidth}
   \caption{NOAA~9866, 14~March 2001.
(a) Second time interval of the time series (the trend has been
subtracted) of the peak brightness temperature at 17~GHz, (b)
wavelet spectrum of the second time interval of the time series.
The eight-minute oscillations are clearly present during the whole
time interval.} \label{F-AR9866}
\end{figure}

It is interesting to note the presence of powerful well-pronounced
eight-minute oscillations~(Figure~\ref{F-AR9866}). The
eight-minute oscillations are less studied than the three- and
five-minute oscillations, although they were observed earlier in
the radio and optical ranges~\cite{Abramov11}. Some authors have
found periodicity of about eight minutes in the 171~\AA\ and
195~\AA\ bandpasses in the coronal loops~(King~\textit{et al.},
2003; Van Doorsselaere, Birtill, and Evans, 2009).
%\cite{King03,Doorsselaere09}.

\subsection{The Active Region 10139 on 7 October 2002}
\label{S-AR10139}

\begin{figure}
   \centerline{\hspace*{0.05\textwidth}
               \includegraphics[width=0.388\textwidth,clip=]{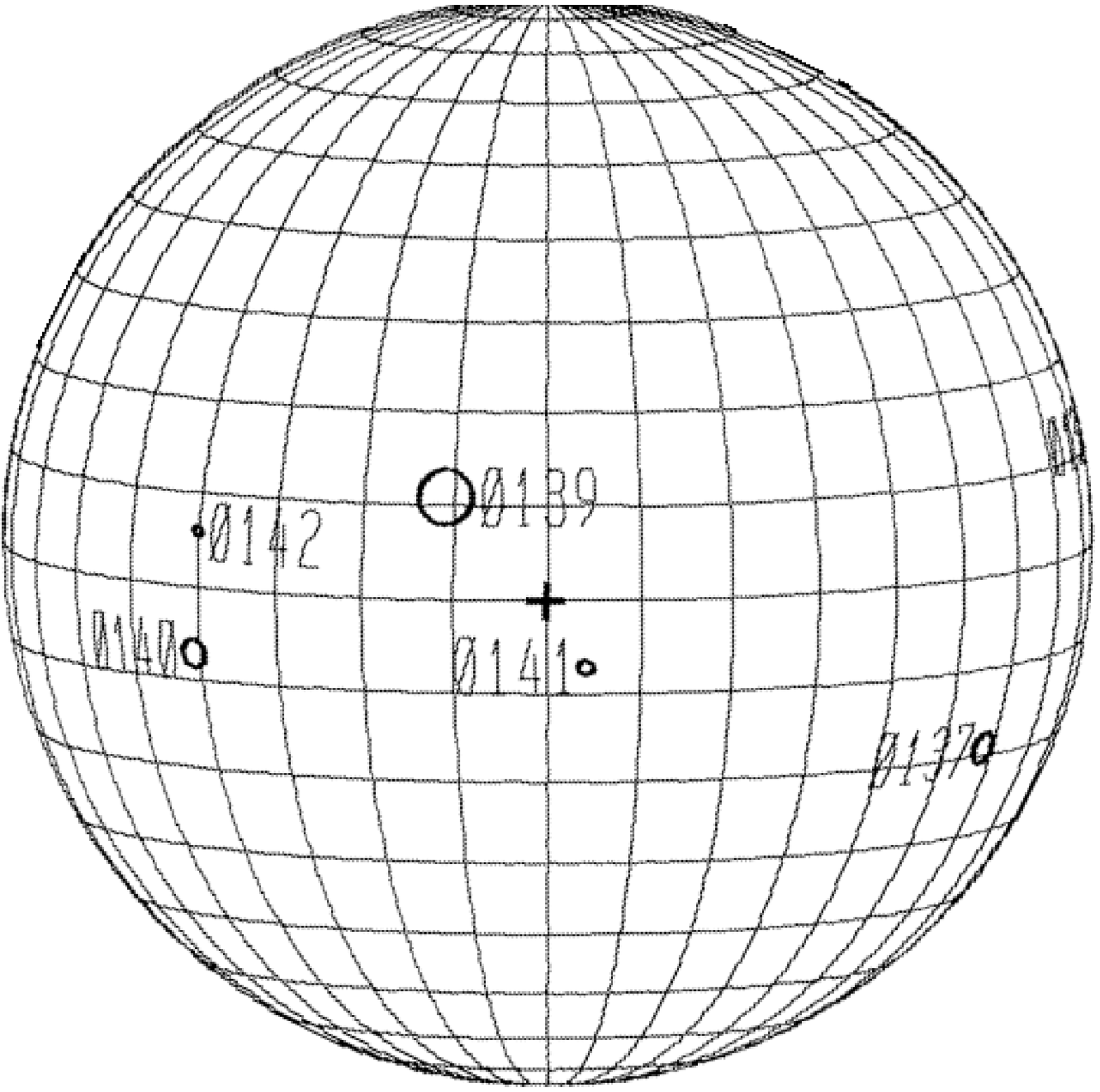}
               \hspace*{0.1\textwidth}
               \includegraphics[width=0.554\textwidth,clip=]{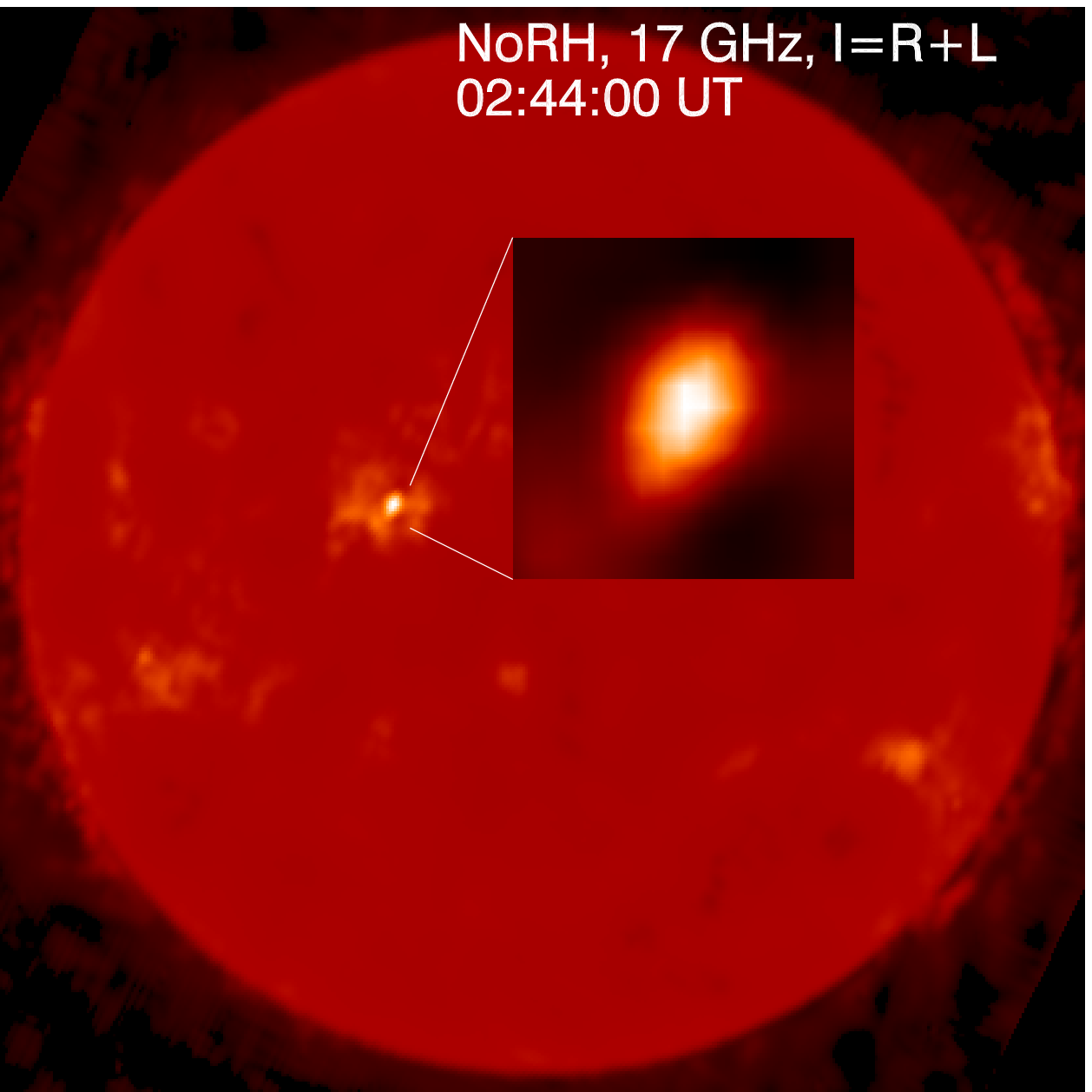}}
   \vspace{-0.38\textwidth}
   \centerline{\Large \bf
               \hspace{0.01\textwidth}\color{black}{a}
               \hspace{0.45\textwidth} \color{white}{b}
   \hfill}
   \vspace{0.38\textwidth}
   \vspace{0.02\textwidth}
   \centerline{\hspace*{-0.05\textwidth}
               \includegraphics[width=0.6\textwidth,clip=]{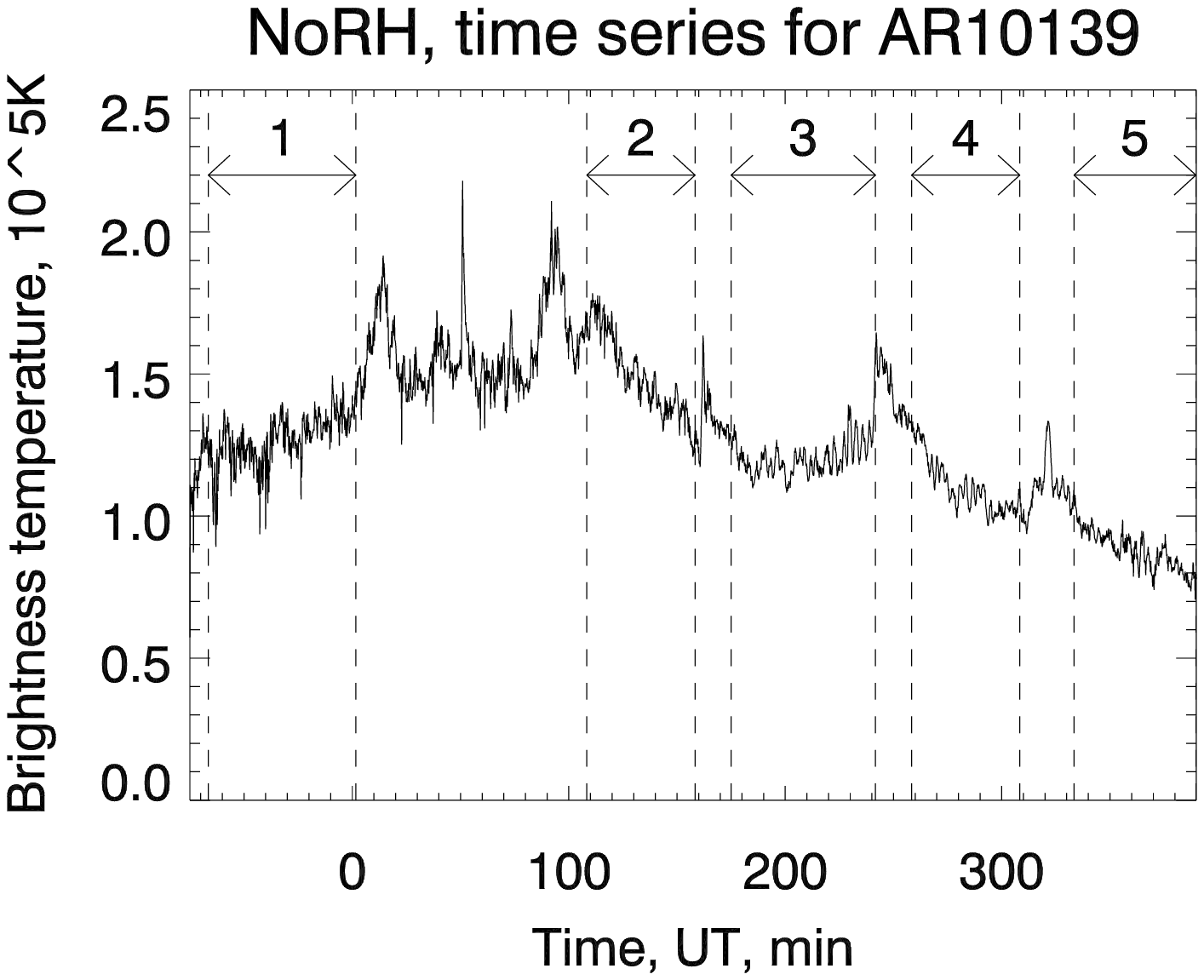}
               \hspace{-0.15\textwidth}
               \includegraphics[width=0.6\textwidth,clip=]{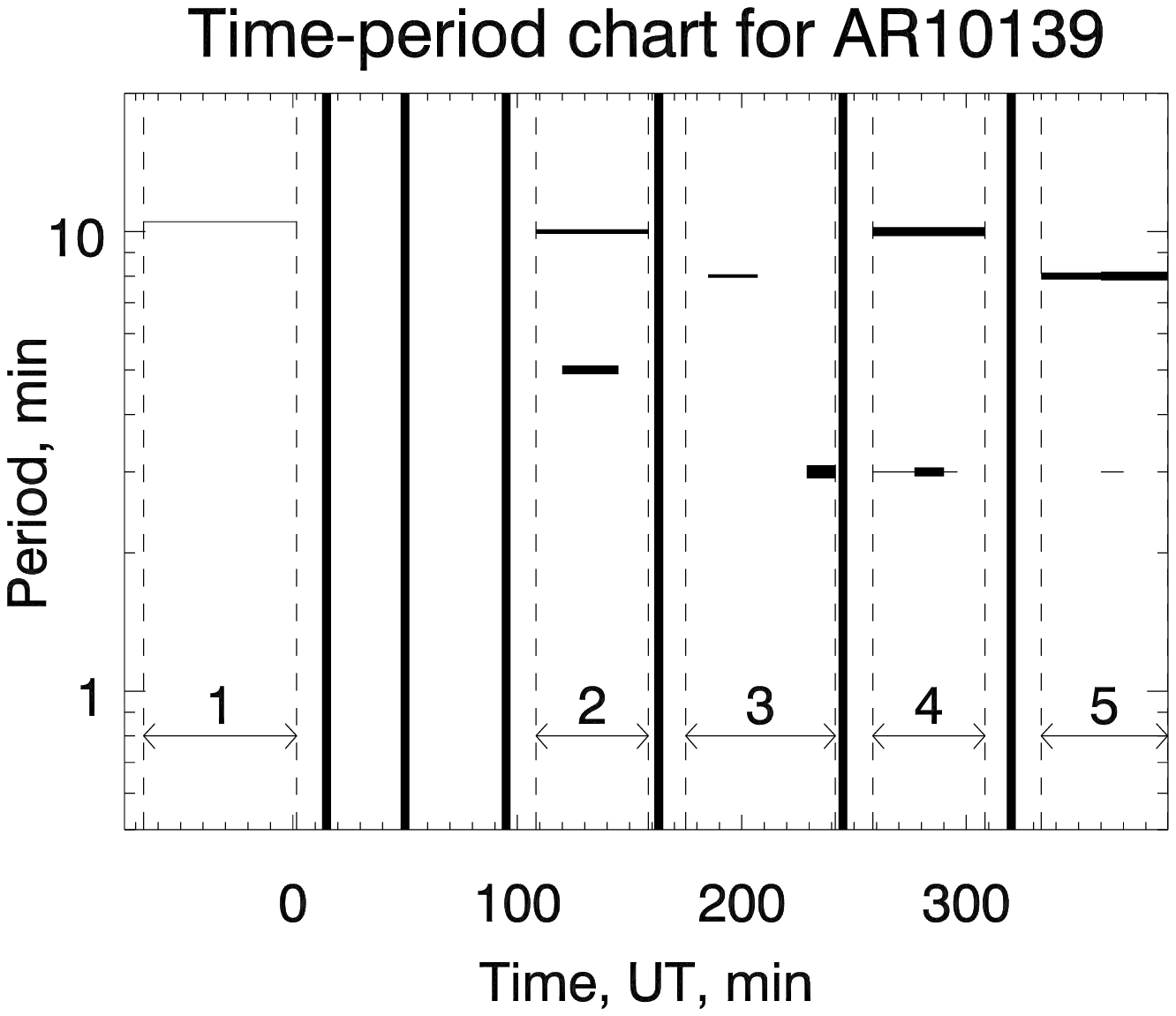}}
   \vspace{-0.46\textwidth}
   \centerline{\Large \bf
               \hspace{0.03\textwidth} \color{black}{(c)}
               \hspace{0.4\textwidth} \color{black}{(d)}
   \hfill}
   \vspace{0.46\textwidth}
\caption{NOAA~10139, 7~October 2002. Top: (a) active region
position on the solar disk (Mees Solar Observatory, 16:32~UT), (b)
radio map at 17~GHz (2:44~UT) and enlarged ROI, the size of the
ROI is about $70''\times70''$. Bottom: (c) time series of the peak
brightness temperature at 17~GHz, vertical dashed lines show five
time intervals of the analysis, (d) time-period chart showing the
life-times of the significant oscillations by the horizontal
lines, the thickness of the lines reflects the wavelet power, the
thick vertical lines show the bursts.} \label{F-2002-10-07}
\end{figure}

The observations were carried out one day before  the
central-meridian passage (CMP). Figure~\ref{F-2002-10-07}a shows
the position of the AR on the solar disk for the date of the
analysis. The overall trend of the emission is caused by daily
variations of the synthesized beam.

There were some small bursts during the day.
Figure~\ref{F-2002-10-07}c shows the time series of the peak
brightness temperature at 17~GHz. For our analysis we used five
time intervals from the time series between the bursts. The
positions of the intervals are shown in Figure~\ref{F-2002-10-07}c
by vertical dashed lines. The time-period chart is depicted in
Figure~\ref{F-2002-10-07}d.

We find that the spectra of the oscillations for different time
intervals differ significantly. The first time interval is noisy.
Practically, there are no oscillations visible. There might be
oscillations with periods about ten minutes or more but we do not
see the three- and five-minute oscillations.

After the bursts, the oscillations appear. During the second time
interval we can see a wave-train of  five-minute oscillations with
duration about 25 minutes. Ten-minute oscillations are also
present during the whole time interval.

\begin{figure}
   \centerline{\includegraphics[width=0.5\textwidth,clip=]{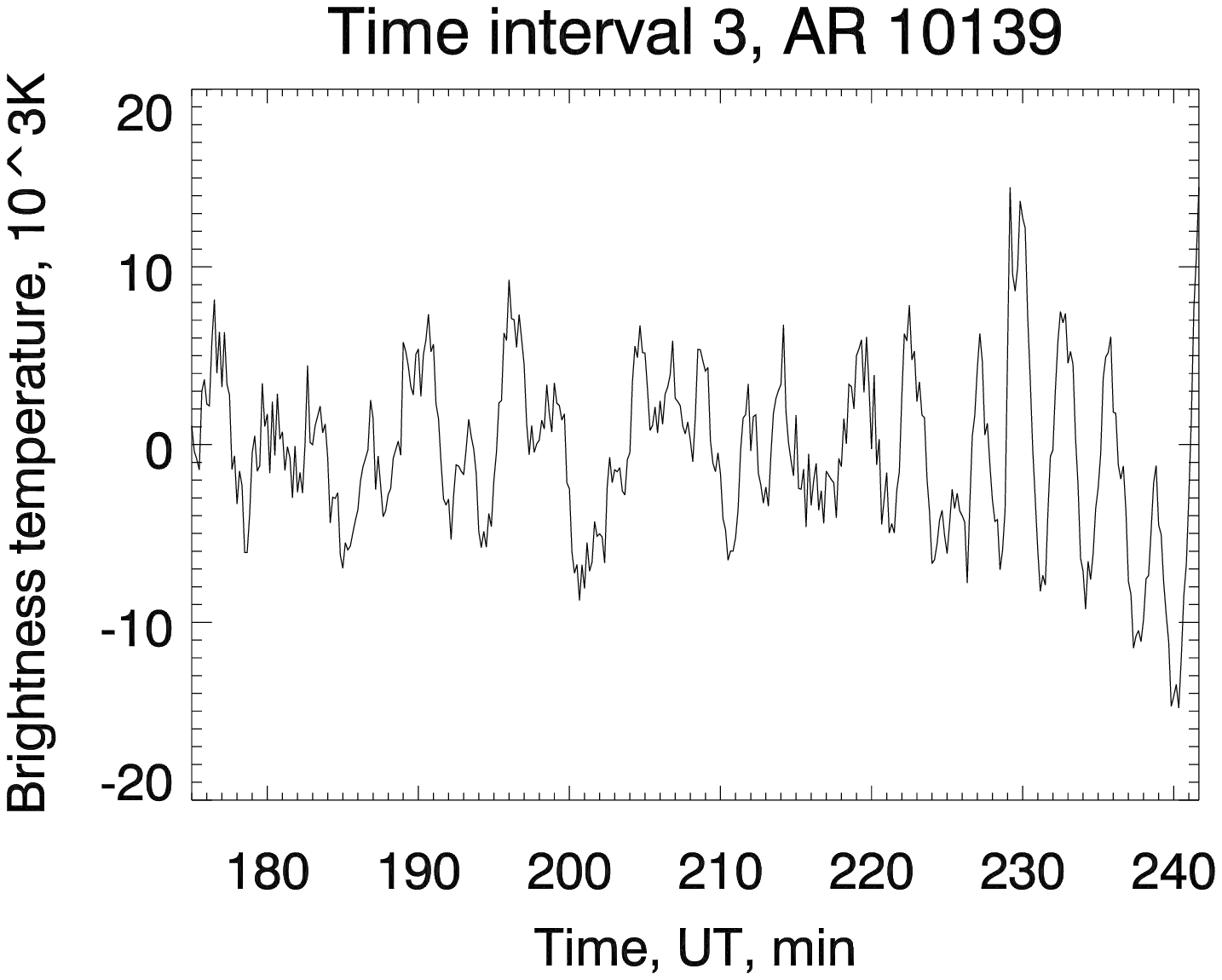}
%               \hspace*{0.1\textwidth}
               \includegraphics[width=0.5\textwidth,clip=]{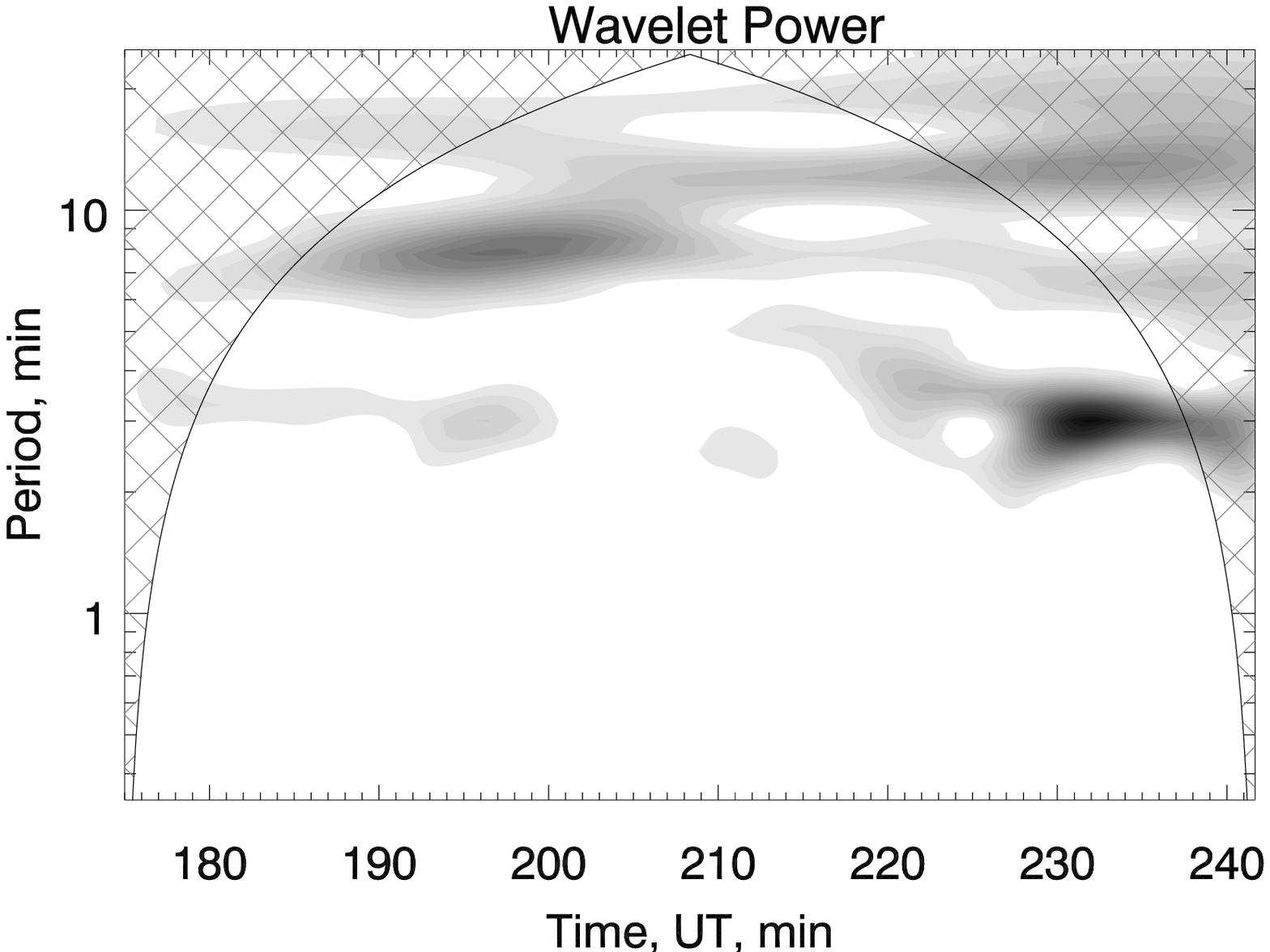}}
   \vspace{-0.35\textwidth}
   \centerline{\Large \bf
               \hspace{0.01\textwidth}\color{black}{a}
               \hspace{0.45\textwidth}\color{black}{b}
   \hfill}
   \vspace{0.35\textwidth}
   \caption{NOAA~10139, 7~October 2002. (a)Third time interval of the time series (the trend has been subtracted)  of the
peak brightness temperature at 17~GHz, (b) wavelet spectra of the
third time interval of the time series. Darker regions correspond
to higher power, the crosshatched area shows the COI. It is clear
visible starting of the well-defined three-minute oscillations at
the end of the interval before the burst.} \label{F-AR10139}
\end{figure}

The third time interval shows the presence of a short wave-train
of  eight-minute oscillations and beginning of  three-minute
oscillations at the end of interval. It is noteworthy that the
well-defined three-minute oscillations begin 15\,--\,20~minutes
before the burst (Figure~\ref{F-AR10139}). Perhaps it is a
precursor, similar to the cases investigated
by~\inlinecite{Sych09}.

In the fourth time interval we can see  three-minute oscillations
and a long wave-train of powerful ten-minute oscillations. In the
fifth interval there is a weak short wave-train of  three-minute
oscillations and a long wave-train of oscillations with periods of
nearly 8\,--\,10~minutes.

\section{Discussion and Conclusions}
\label{S-Discussion}

We have presented three cases that confirm the existence of a link
between oscillation spectrum and flare activity.

The analysis of the wavelet spectra of sunspot oscillations at the
wavelength of 1.76~cm have confirmed the presence of the
variations in the spectra connected with the flare activity. Both
disappearance and appearance of some periods were found.

The first case (NOAA 9608, 11 September 2001) shows a decrease of
power of the eight-minute oscillations after the first burst and
their disappearance after the second, more powerful burst. This
case also shows an increase of power of the three- and five-minute
oscillations between the bursts. The second case (AR 9866, 14
March 2002) shows the appearance of the eight-minute oscillations
after the burst. The most impressive feature in the third case
(NOAA 10139, 7 October 2002) is the beginning of well-defined
three-minute oscillations 15\,--\,20 minutes before the burst.

Two of the three cases demonstrate an increase of power of the
short-term oscillations before and during the bursts. This result
is in a good agreement with the results obtained
by~\inlinecite{Sych09}. They found similar cases of a gradual
increase in the power of the three-minute oscillations before
flares in sunspot associated sources. They proposed that slow
magnetoacoustic waves propagate from a sunspot along coronal loops
upwards the flare site and cause the energy release. This
interpretation is  likely to be valid for our cases, too.

In our previous study~\cite{Abramov11} we showed observational
evidence that waves travelled from the chromosphere towards the
corona inside the umbral magnetic flux tube of the sunspot. We
found similar wave-trains in the oscillation processes at two
different levels of the solar atmosphere and measured time shift
between wave-trains. In this study we confirmed the relationship
between travelling waves and flare activity. One of the possible
interpretation of the presented observational material is that the
MHD waves propagating from the sunspot can trigger energy release
at the flare site.

The  increase of power of the three-minute oscillations before the
burst can be used for flare forecasting. Of course, there are some
difficulties. First, the detection of the precursor requires a
complicated data processing. Second, this effect appears only a
few minutes before the flare. So, further studies  are needed.
Studies of this effect may shed light on the problem of energy
release in solar flares. NoRH is a very suitable instrument for
such investigations due to its long series of observations (since
1992), high time (1~sec) and spatial (10~arcsec) resolution for
the full solar disk, and  its long daily observing time (up to
8~hours per day).

Our conclusions are as follows:

1. We have confirmed the existence of a link between the
oscillation spectrum and flare activity.

2. We found one case of increase of the power of the short-term
oscillations before the burst.

3. We interpret our observations in terms of magnetohydrodynamic
waves propagating from sunspots and flare processes.

4. The NoRH is a very suitable instrument for such studies.

%%%%%%%%%%%%%%%%%%%%%%%%%%%%%%%%%%%%%%%%%%%%%%%%%%%%%%%%%%%%%%%%%%%%%%%%%%%
\begin{acks}
This work was partially supported  by the Presidium of Russian
Acade\-my of Sciences under grant OFN-15.

We are grateful to the anonymous referee and Editors for important
remarks, which helped us to improve the paper.

Data used here from Mees Solar Observatory, University of Hawaii,
are produced with the support of NASA grant NNG06GE13G.

Wavelet software was provided by C.~Torrence and G.~Compo, and is
available at URL: http://paos.colorado.edu/research/wavelets/
\end{acks}

\end{article}
\end{document}